# Adaptive synchronization of coupled chaotic oscillators

Bhargava Ravoori[1,2], Adam B. Cohen[1,2], Anurag V. Setty[1,3], Francesco Sorrentino[1,4], Thomas E. Murphy[1,5], Edward Ott[1,2,5], Rajarshi Roy[1,2,3]

[1]*Institute for Research in Electronics and Applied Physics, University of Maryland, College Park, Maryland 20742, USA*

[2]*Department of Physics, University of Maryland, College Park, Maryland 20742, USA*

[3]*Institute for Physical Science and Technology, University of Maryland, College Park, Maryland 20742, USA*

[4]*Università degli Studi di Napoli Parthenope, 80143 Napoli, Italy*

[5]*Department of Electrical and Computer Engineering, University of Maryland, College Park, Maryland 20742, USA*

29[th] September, 2009

We experimentally demonstrate and numerically simulate a new adaptive method to maintain synchronization between coupled nonlinear chaotic oscillators, when the coupling between the systems is unknown and time-varying (e.g., due to environmental parameter drift). The technique is applied to optoelectronic feedback loops exhibiting high dimensional chaotic dynamics. In addition to keeping the two systems isochronally synchronized in the presence of *a priori* unknown time-varying coupling strength, the technique provides an estimate of the time-varying coupling.

## I. Introduction

A surprising result of past research [1-4] is that, even if several individual systems behave chaotically, in the case where the systems are identical (near identical), by proper coupling, the systems can be made to evolve toward a situation of exact (approximate) isochronal synchronism. That is, in the case of identical systems, the chaotic orbits of the individual systems are precisely the same, and at every instant of time the states of each of the coupled systems are all equal. Moreover, the coupling can be designed so that, in the synchronized state, the coupling terms become zero and are only nonzero when the states of the individual coupled systems are out of synchrony [5]. Various uses of this phenomenon have been proposed. These include secure communication [2, 6], use of the chaotic signal to encode information [7], parameter estimation and prediction [8, 9] and sensors [10]. In all of these applications, synchronism of chaos is critically dependent upon maintenance of a proper coupling between the systems. In general, coupling may be thought of as utilizing channels through which the individual systems exchange partial information about their states. In this paper we consider the case where these channels are subject to time dependent changes that are unknown to the individual coupled systems, and we address the issue of how to adaptively adjust the coupling to maintain synchrony in the presence of such changes. We emphasize that channel change or 'drift' is a potentially very common situation (e.g., due to environmental changes that affect the channels) and its effect may be crucial to the determination of the robustness of the above mentioned applications.

Adaptive synchronization strategies in a network of nonlinear oscillators have been a topic of much interest [e.g., 10, 11, 12]. Sorrentino and Ott [10] proposed and simulated an efficient adaptive algorithm that maintains synchronization among a network of connected dynamical systems in the presence of *a priori* unknown time-varying coupling. In this work, we experimentally demonstrate this scheme using a pair of nonlinear time-delayed optoelectronic feedback loops as the dynamical systems. These systems are similar to those used recently for chaotic communication by Kouomou *et al.* [13] and Argyris *et al.* [6]. The technique is shown to both maintain isochronal synchronization [14, 15] between the two systems and, in the process, generate a real-time estimate of the originally unknown time variations of the coupling.

**II. Experimental setup**

Each feedback loop in our experimental setup (Fig. 1) comprises a semiconductor laser which serves as the optical source, a Mach-Zehnder electro-optic modulator, a photoreceiver, an electronic filter and an amplifier. The electronic filter is chosen to be a two-pole bandpass filter. The dynamics of the feedback loop can be modeled by the delay-differential equations [8, 13],

$$\omega_L^{-1} \frac{dx_1(t)}{dt} = -\left(1 + \frac{\omega_H}{\omega_L}\right) x_1(t) - y_1(t) - \beta \cos^2[x_1(t-\tau) + \varphi_o],$$

$$\omega_H^{-1} \frac{dy_1(t)}{dt} = x_1(t).$$

(1)

Here $x_1(t)$ is the normalized voltage signal applied to the electro-optic modulator, $\tau$ is the feedback time delay and $\varphi_o$ is the bias point of the modulator. $\omega_L$ and $\omega_H$ are the filter's lowpass and highpass corner frequencies. $\beta$ is the dimensionless feedback strength proportional to the optical power $P_o$ entering the electro-optic modulator. The optical output of the electro-optic modulator is $P_1(t) = P_o \cos^2(x_1(t) + \varphi_o)$. Incorporating a programmable component as part of the feedback loop allows us to perform real-time computations such as the implementation of an adaptive synchronization algorithm. We use a digital signal processor (DSP) based programmable board (Spectrum Digital DSK6713) to perform electronic filtering and time delay. The collected samples are filtered digitally using the DSP. The filtered signal is then stored in memory for a desired amount of time before being output. This, combined with processor latency, produces the total loop time delay $\tau$. The discrete time difference equations describing the digital filter are presented and discussed in [16].

Depending on the value of the feedback strength and delay, the loop is capable of producing dynamics ranging from periodic oscillations to high-dimensional chaos [8, 13]. For the measurements reported here, we adjusted the laser power $P_o$ to yield a feedback strength of $\beta = $ 3.58, and we programmed the DSP board to produce a net feedback delay of 1.5 ms. The highpass cut-on frequency was designed to be $\omega_H/(2\pi) = 100$ Hz, and the lowpass cut-off frequency was chosen to be $\omega_L/(2\pi) = 2.5$ kHz. The bias point of the electro-optic modulator is adjusted to $\varphi_o = \pi/4$. With this choice of parameters, the system exhibits high-dimensional chaotic dynamics.

We couple two nominally identical optoelectronic feedback loops uni-directionally, *i.e.*, the transmitter affects the dynamics of the receiver, but not vice versa. Thus the equations of motion describing the coupled system are given by Eq. (1) for the transmitter and

$$\omega_L^{-1} \frac{dx_2(t)}{dt} = -\left(1 + \frac{\omega_H}{\omega_L}\right) x_2(t) - y_2(t) - \beta \cos^2[(1 - \bar{\kappa}(t))x_2(t-\tau) + r(t) + \varphi_o],$$

$$\omega_H^{-1} \frac{dy_2(t)}{dt} = x_2(t), \quad (2)$$

$$r(t) = \kappa(t) x_1(t-\tau),$$

for the receiver. In Eqs. (2), $\kappa$ denotes the coupling strength. It is envisioned that the signal received by system 2 (Eqs. (2)) is the product $r(t) = \kappa(t) x_1(t-\tau)$, but that system 2 has no knowledge of $\kappa(t)$ and $x_1(t-\tau)$ individually. System 2, however, has control of the quantity $\bar{\kappa}(t)$. Furthermore, we note from Eqs. (2) that the coupled system admits a synchronous solution if $\kappa = \bar{\kappa}$. For the case of time independent $\kappa = \kappa_o$, numerical simulations and experimental observations show that the synchronous solution is *stable* if $0.45 < \kappa_o < 1.49$. Thus if the receiver quantity $\bar{\kappa}(t)$ can correctly match $\kappa(t)$, and if the latter varied sufficiently slowly in time, then the system can be maintained in a synchronous state through time variation of $\kappa$ over the above values.

In our experiments the optical output of the electro-optic modulator $P_1$ is transmitted to the receiver loop through a fiber-optic link. The received optical signal is detected and filtered at the receiver to build $x_1(t-\tau)$. The communication channel is subject to simulated time variation modeled by a time-varying coupling strength $\kappa = \kappa(t)$ so that the received signal used in the coupling term of Eq. (2) is $r(t) = \kappa(t)x_1(t-\tau)$. Here we assume the variation of the coupling strength occurs very slowly compared to the dynamics of each feedback loop, *i.e.*, the time scale for variation of $\kappa(t)$ is substantially longer than that for $x_{1,2}$. Our goal is to use the available information $r(t)$, to dynamically construct an estimate of the coupling $\kappa(t)$ by adaptively changing $\bar{\kappa}(t)$ in a manner which maintains synchrony.

## III. Adaptive strategy

We implement the adaptive strategy proposed by Sorrentino and Ott [10]. The scheme relies on the minimization of a 'potential' $\psi$, defined as the exponentially weighted moving average of the synchronization error,

$$\psi(t) = \frac{1}{\tau_o} \int_{-\infty}^{t} (\bar{\kappa}(t')x_2(t'-\tau) - r(t'))^2 \exp[-(t-t')/\tau_o] \, dt' \,. \tag{3}$$

Minimizing (3) with respect to $\bar{\kappa}$, we arrive at the following equations which are then solved to construct of the estimate $\bar{\kappa}(t) = N(t)/D(t)$ of the channel time-variation $\kappa(t)$, where $N(t)$ and $D(t)$ follow

$$\begin{aligned}\tau_o \frac{dN(t)}{dt} + N(t) &= r(t)x_2(t-\tau), \\ \tau_o \frac{dD(t)}{dt} + D(t) &= x_2^{\,2}(t-\tau).\end{aligned} \tag{4}$$

Note that $N(t)$ and $D(t)$ in Eqs. (4) are simply the lowpass filtered versions of the 'signals' on the right hand sides of these equations and $\tau_o$ denotes the time constant of the filter. For the success of the adaptive algorithm it is important to choose $\tau_o$ to be large compared to the dynamical time scale of the chaotic oscillators but small compared to the time scale on which $\kappa(t)$ varies. In our experiments and simulations we choose $\tau_o = 0.81$ ms.

**IV. Results and discussion**

The results from the application of the adaptive synchronization technique are presented in Figure 2. We show the received optical signal $\kappa(t)P_1(t)$ and the difference error signal $(P_1 - P_2)$ both normalized to the input optical power $P_o$. The received signal reflects the long time-scale coupling strength variations as a result of the simulated environmental time fluctuations in $\kappa(t)$. The time variation of $\kappa(t)$ was accomplished by inserting into the coupling channel an electro-optic intensity modulator driven by an arbitrary waveform generator. The coupling strength $\kappa(t)$ varies between 0.83 and 1.13 with an average value of 0.98 and a repetition period of 500 ms. Before $t = 0$, the adaptive synchronization technique is disabled and $\bar{\kappa}$ is held fixed at 0.8. During this period, the synchrony between the two feedback loops is poor, as shown by the difference signal $(P_1 - P_2)$. After $t = 0$, the adaptive algorithm is switched on. Now, the receiver forms a real-time estimate of the time dependent coupling and dynamically compensates to maintain synchrony. The difference signal is seen to reduce in amplitude showing successful synchronization of the receiver to the transmitter dynamics. The estimate of the channel variation $\bar{\kappa}(t)$ is also shown along with the channel variations $\kappa(t)$. Corresponding results from numerical simulations are also presented. Figure 3 shows two 10 ms time windows, one where the adaptive scheme is off ($-440$ ms $< t < -430$ ms) and one where the adaptive scheme is on (60 ms $< t <$ 70 ms), to detail the dynamics of the feedback loops occurring at sub-millisecond time scales.

The adaptive synchronization technique successfully tracks slowly varying coupling strength time variations and compensates for them to maintain synchrony between the transmitter and the receiver systems. To evaluate the efficiency of the adaptive algorithm, we impose a coupling strength time-variation of the form $\kappa(t) = 0.8\,(1+\varepsilon \sin(2\pi\, f_{\text{mod}}\, t))$ and measure a normalized synchronization error, $\sigma = \sqrt{\langle (P_1 - P_2)^2 \rangle}/P_o$, for different values of $f_{\text{mod}}$. Figure 4(a) shows the experimentally measured synchronization error for a nominal modulation depth $\varepsilon = 0.4$, as a function of the modulation frequency, comparing the case when the adaptive strategy is enabled (solid line) to when it is disabled (dotted line). At all frequencies observed, the adaptive technique yields an improvement in the degree of synchronization compared to the uncontrolled case, with the greatest improvement seen at lower frequencies. Numerical simulations (lower dashed line) of the adaptive synchronization method show the theoretically possible improvement in synchronization. These simulations do not include noise and parameter mismatches; incorporating these non-ideal effects (not shown) allows us to improve the correspondence between the simulation results and experimental observations. We have also found that with the random channel variations we have simulated, a simple envelope detection scheme with a lowpass filter to estimate $\overline{\kappa}(t)$ fails to produce synchrony.

The degree to which the receiver loop tracks coupling strength variations is quantified by a tracking measure, $\mu = (\tilde{\overline{\kappa}}(f_{\text{mod}})/\tilde{\kappa}(f_{\text{mod}}))$, defined as the ratio of the Fourier amplitudes at the frequency $f_{mod}$ of the tracking estimate $\bar{\kappa}$ to that of $\kappa$. Figure 4(b) shows the experimentally measured tracking measure as a function of the channel time-variation frequency $f_{mod}$ obtained using a vector network analyzer (Agilent 4395A). Results are shown for various modulation depths, $\varepsilon = 0.1, 0.2, 0.4$. For large modulations, the tracking measure is seen to start deteriorating at a lower frequency compared to the case when the modulation is small.

To summarize, the synchronization of chaotic systems sensitively depends on the coupling strength which may be perturbed by environmental changes. We experimentally demonstrate an adaptive algorithm to maintain synchrony between chaotic systems when the coupling channel variations are unknown but slow in comparison to the chaotic dynamics. In addition to keeping the chaotic systems synchronized, the scheme also allows us to estimate the time-varying coupling strength. We note that by using faster electronics, the system can easily be scaled to allow the adaptive scheme to track faster channel variations.

**Acknowledgements**

We thank Karl Schmitt and Caitlin Williams for advice and help. This work is supported by DOD MURI grant (ONR N000140710734) and the US-Israel Binational Science Foundation.

**References**

[1] H.Fujisaka and Y. Yamada, Prog. Theor. Phys. **69**, 32 (1983); V. S. Afraimovich, N. N. Verichev, and M. I. Rabinovich, Izvest. Vys. Uch. Zav. Radiofizika **29**, 1050 (1986); L. M. Pecora and T. L. Carroll, Phys. Rev. Lett. **64**, 821 (1990), L. M. Pecora and T. L. Carroll, Phys. Rev. Lett. **80**, 2109 (1998).

[2] K. M. Cuomo and A. V. Oppenheim, Phys. Rev. Lett. **71**, 65 (1993).

[3] G. D. VanWiggeren and R. Roy, Science **279**, 1198 (1998); J. P. Goedgebuer, L. Larger and H. Porte, Phys. Rev. Lett. **80**, 2249 (1998);

[4] A. Arenas, A. Díaz-Guilera, J. Kurths, Y. Moreno, and C. Zhou, Physics Reports **469**, Issue 3, 93-153 (2008).

[5] A. Pikovsky, M. Rosenblum, and J. Kurths, *Synchronization: A Universal Concept in Nonlinear Sciences (Cambridge Nonlinear Science Series)*, Cambridge University Press (2003).

[6] A. Argyris et al., Nature **438**, 343 (2005).

[7] V. Dronov, M. R. Hendrey, T. M. Antonsen, Jr, and E. Ott, Chaos **14**, 30 (2004).

[8] A.B. Cohen, B. Ravoori, T. E. Murphy, and R. Roy, Phys. Rev. Lett. **101**, 154102 (2008).

[9] J. C. Quinn, P. H. Bryant, D. R. Creveling, S. R. Klein, and H. D. I. Abarbanel, arXiv:0904.2798 (2009).

[10] F. Sorrentino and E. Ott, Phys. Rev. Lett. **100**, 114101 (2008); F. Sorrentino and E. Ott, Phys. Rev. E **79**, 016201 (2009).

[11] C. Zhou and J. Kurths, Phys. Rev. Lett. **96**, 164102 (2006).

[12] J. Ito and K. Kaneko, Phys. Rev. Lett. 88, 028701 (2001); F. Mossayebi, H. K. Qammar, and T. T. Hartley, Phys. Lett. A **161**, 255 (1991); C. R. Johnson, Jr. and J. S. Thorp, IEEE Sig. Proc. Lett. **1**, 194 (1994); D. J. Sobiski and J. S. Thorp, IEEE Trans. Circ. Syst. I **45**, 194 (1994);

U. Parlitz and L. Kocarev, Int. J. Bifurcation Chaos Appl. Sci. Eng. **6**, 581 (1996); C. W. Wu, T. Yang and L. O. Chua, Int. J. Bifurcation Chaos Appl. Sci. Eng. **6**, 455 (1996); L. O. Chua, T. Yang, G.-Q. Zhong, and C. W. Wu, Int. J. Bifurcation Chaos Appl. Sci. Eng. **6**, 189 (1996).

[13] Y. C. Kouomou, P. Colet, L. Larger, and N. Gastaud, Phys. Rev. Lett. **95**, 203903 (2005).

[14] I. Fischer *et al.*, Phys. Rev. Lett. **97**, 123902 (2006).

[15] B. B. Zhou and R. Roy, Phys. Rev. E **75**, 026205 (2007).

[16] T. E. Murphy *et al.* (unpublished), Phil. Trans. R. Soc. A. (2009)

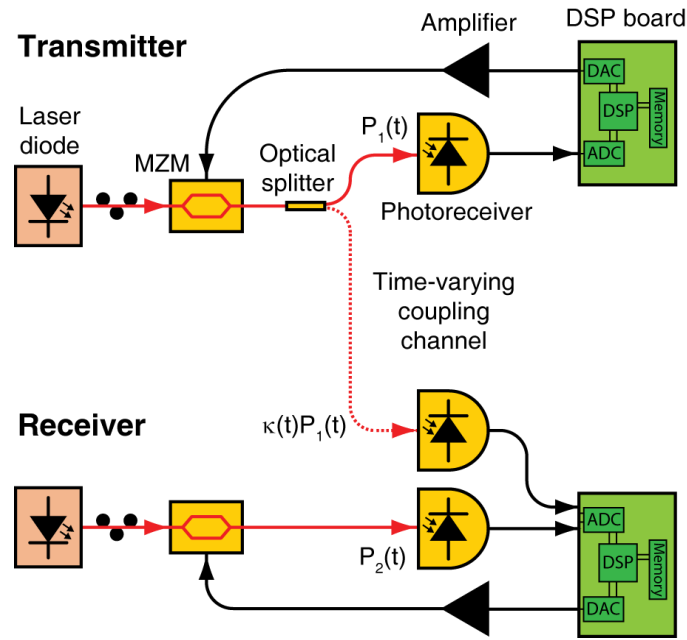

FIG. 1 (color online). The transmitter and receiver systems are optoelectronic feedback loops coupled through an optical channel that may be subject to environmental drift in time, thus perturbing the coupling strength between the two dynamical systems. The DSP board contains a dual input analog-to-digital converter (ADC), a signal processing unit, memory, and a digital-to-analog converter (DAC).

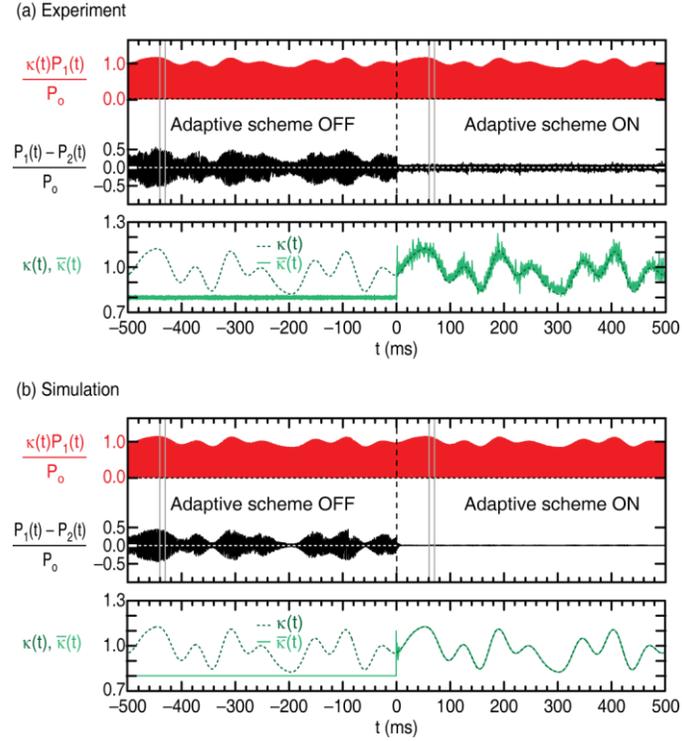

FIG. 2 (color online). Success of adaptive strategy for tracking and synchronization in the presence of a time-varying coupling strength. Before time $t = 0$, the adaptive synchronization algorithm is disabled. During this period, $\kappa(t)$ fluctuates between 0.83 and 1.13 while the coupling strength estimate $\bar{\kappa}$ is held constant at 0.8. After $t = 0$ (indicated by the dashed vertical line), the adaptive algorithm is turned on and a real-time estimate of the channel modulation is constructed from measurements performed entirely at the receiver. (a) Results from experiment. The top panel shows the received optical signal $\kappa(t)P_1(t)$ and the error signal $(P_1(t) - P_2(t))$, both normalized to the optical power $P_o$ entering the electro-optic modulator. The bottom panel shows the time dependence of the coupling strength $\kappa(t)$ and its real-time estimate $\bar{\kappa}(t)$ obtained once the adaptive algorithm is switched on. (b) Corresponding results from numerical simulations. The solid vertical lines indicate the four intervals of time we present in more detail in Fig. 3.

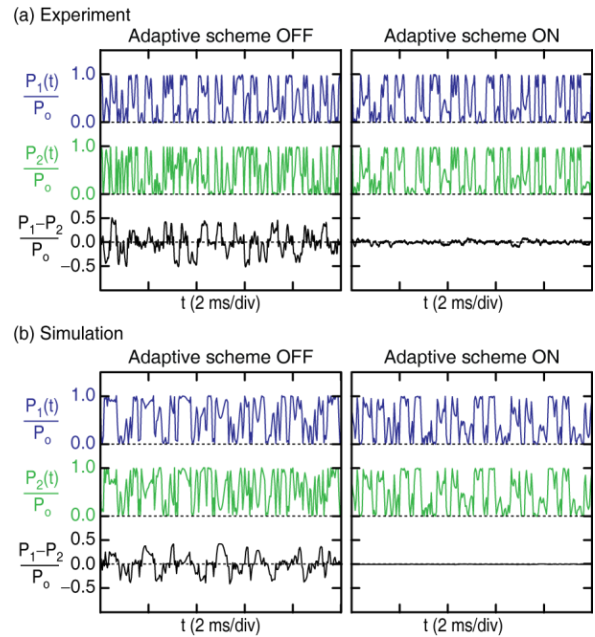

FIG. 3 (color online). Synchronization of transmitter and receiver dynamics using adaptive scheme. (a) The left and right panels are 10 ms time traces of the internal dynamics $P_1(t)$ and $P_2(t)$ and the error signal $(P_1(t) - P_2(t))$ corresponding to the windows in Fig. 2(a). Under the same coupling conditions, the two feedback loops only synchronize when the adaptive strategy is enabled. (b) Results from simulations corresponding to the windows in Fig. 2(b).

14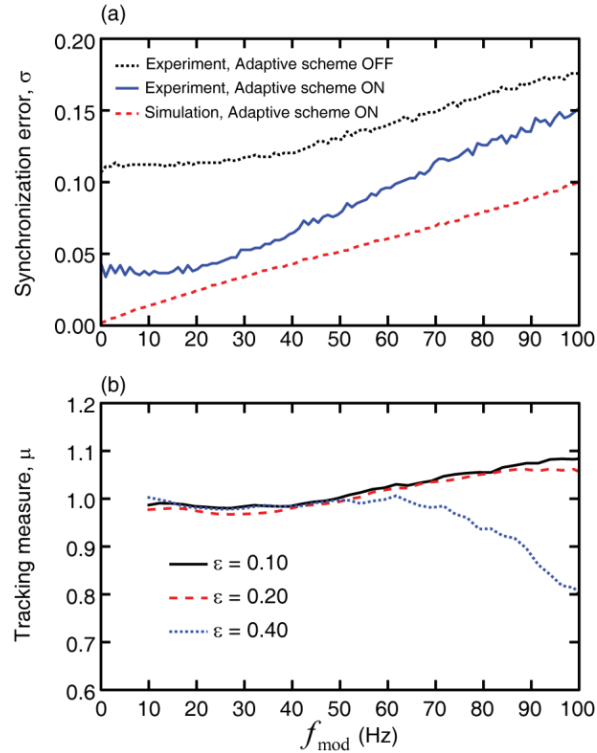

FIG. 4 (color online). (a) Synchronization error $\sigma$ as a function of modulation frequency $f_{mod}$ for a modulation depth $\varepsilon = 0.40$. The experimentally measured synchronization error when the adaptive strategy is ON (solid line) is lower than when the adaptive strategy is OFF (dotted line). Numerical simulations (dashed line) result in a smaller $\sigma$ compared to the experimental observations. (b) Tracking measure $\mu$ is defined as the ratio of the Fourier amplitude of the receiver's tracking estimate $\bar{\kappa}(t)$ at frequency $f_{mod}$ to the Fourier amplitude of an imposed channel modulation. Experimental tracking measure for modulation amplitudes $\varepsilon = 0.1, 0.2,$ and 0.4 are shown.

14